\documentclass[preprintnumbers,eqsecnum,twocolumn,prd]{revtex4}  %showpacs,showkeys,preprint

\usepackage{graphicx}
\usepackage{amsmath}

\def\beq{\begin{equation}}
\def\eeq{\end{equation}}
\def\rmd{{\rm d}}

\def\rdotl{(\boldsymbol{r}\cdot\boldsymbol{\bar{\ell}}_{\not\,0})}
\def\robsdotl{(\boldsymbol{r}_{\rm obs}\cdot\boldsymbol{\bar{\ell}}_{\not\,0})}
\def\vdotl{(\tilde{\boldsymbol{v}}\cdot\boldsymbol{\bar{\ell}}_{\not\,0})}
\def\vdotb{(\tilde{\boldsymbol{v}}\cdot\boldsymbol{b})}
\begin{document}

\title{Gravitational wave effects on astrometric observables}

\author{Donato \surname{Bini}$^{1,2,3}$}
\author{Andrea \surname{Geralico}$^1$}

\affiliation{
$^1$Istituto per le Applicazioni del Calcolo ``M. Picone,'' CNR, I--00185 Rome, Italy\\
$^2$INFN Sezione di Napoli, Complesso Universitario di Monte S. Angelo,
Via Cintia Edificio 6, 80126 Naples, Italy\\
$^3$ International Center for Relativistic Astrophysics Network, I--65122 Pescara, Italy
}

\date{\today}

\begin{abstract}
Observational data from the ESA astrometric mission Gaia determining the positions of celestial objects within an accuracy of few microarcseconds will be soon fully available.
Other satellite-based space missions are currently planned to significantly improve such precision in the next years.
The data reduction process needs high-precision general relativistic models, allowing one to solve the inverse ray-tracing problem in the gravitational field of the Solar System up to the requested level of accuracy and leading then to the estimate of astrometric parameters. 
Besides a satisfactory description of the background field due to the planets (which should include their multipolar structure), one should consider also other effects which may induce modifications to the light propagation. For instance, the interaction of the light signal with the superposed gravitational field of a gravitational wave emitted by a distant source would cause a shift in the apparent positions of the stars. 
We compute here the main astrometric observables needed for data reduction of satellite-based missions in the presence of a passing plane gravitational wave. We also take into account the effect of the mass quadrupole moment of the planets, improving previous results obtained for Gaia.
\end{abstract}

\pacs{04.20.Cv}

\maketitle

\section{Introduction}

The aim of modern astrometry is to determine, with very high accuracy, the position and proper motion of the stars from satellite-based angular observations.
The ESA mission Gaia launched in 2013 is expected to produce a star catalog within an accuracy of few microarcseconds \cite{Gaia:2016,gaiaweb}, but future space missions should reach a precision of sub-microarcseconds or even nanoarcseconds (see, e.g., Refs. \cite{neat,astrod,lator,odyssey}).
The fully general relativistic modelling necessary to locate a celestial object with such an accuracy requires a detailed account of the underlying measurement process as well as a likewise accurate description of the background gravitational field.
The baseline model for the Gaia data reduction is called the Gaia relativistic model (GREM) \cite{Klioner:1991,Klioner:1992,Klioner:2003,Klioner:2003zi,Klioner:2003ss,Klioner:2003ew,Klioner:2010pu,Zschocke:2010qn}.
This model has been recently improved in Refs. \cite{Zschocke:2015mca,Zschocke:2016rng}, where the light propagation in the gravitational field of $N$ arbitrarily moving bodies of finite size has been determined in the first post-Newtonian (PN) and in the 1.5PN approximation, respectively. 
A different model called the relativistic astrometric model (RAMOD) has been formulated in Refs. \cite{deFelice:2004nf,deFelice:2006xm} (see also Ref. \cite{Crosta:2015jia} and references therein for further developments). 
The astrometric observables associated with Gaia have been recently computed in Ref. \cite{Crosta:2017eco} in the case of pointlike sources moving with constant velocities.

It has also been suggested to use high-precision astrometry to investigate the shift in the apparent positions of the stars induced by gravitational waves (GWs) \cite{Braginsky:1989pv,Pyne:1995iy}.
The observed angular deflections are expected to be of the order of the characteristic strain amplitude of the wave and to have a characteristic pattern, so that measuring them would allow for an indirect detection of the GW itself. 
Various kinds of gravitational waves in this context have been discussed so far, including a stochastic GW background \cite{Book:2010pf} and gravitational waves from localized sources \cite{Kopeikin:1998ts}.
Gaia observes more than a billion stars over an operating period of 5--10 years, locating each of them about 80 times (in 5 years). The sensitivity bandwidth of Gaia to GWs is estimated between $10^{-9}$ Hz and $10^{-5}$ Hz \cite{Klioner:2017asb}. 
A technique to search for low-frequency GWs in the Gaia dataset has been proposed in Ref. \cite{Moore:2017ity}, where it has been also tested in the case of a simulated GW event produced by a supermassive black hole binary system on a circular orbit.
This method is complementary to the pulsar timing approach, which uses the precise timing of millisecond pulsars to search for low-frequency GWs and measure their polarization. The use of astrometry to constrain the polarization content of GWs has been discussed in Refs. \cite{Mihaylov:2018uqm,OBeirne:2018slh}.
Astrometric signatures of GWs can also be found in the residuals of the astrometric solution (which takes into account the deflection of light due
to solar system bodies only), if the period of the GW is much smaller than the time span of the data, as discussed in Ref. \cite{Klioner:2017asb} in the case of a plane wave. 
The parameters characterizing the GW will enter the astrometric model together with the source parameters (i.e., position, proper motion and parallax) and satellite parameters (attitude and calibration), but require a proper estimation process to be determined.

The aim of the present work is twofold. We will study the effect of a passing gravitational plane wave to the main observables of a satellite-based astrometric mission including Gaia, i.e., to the direction cosines measured by the observer on the satellite which are related to the along-scan and across-scan measurements.
Furthermore, we will extend the results of Ref. \cite{Crosta:2017eco} by including the effect of the quadrupole moment of the planets, which cannot be neglected for astrometry at the microarcsecond level of accuracy of Gaia, as discussed in Refs. \cite{Klioner:1991,Klioner:2003,Zschocke:2010qn,Zschocke:2015mca}.

We follow here standard notations and conventions: units are chosen so that $c=1=G$, but they are restored when necessary; the metric signature is $+2$; Greek indices run from 0 to 3, whereas Latin indices run from 1 to 3; unless differently specified, scalar product operation is defined with respect to the spacetime metric.

\section{Setup of the problem}

\subsection{Coordinate choice and background spacetime metric}
 
The background spacetime consists of $N$ gravitationally interacting bodies, each associated with its own world tube (approximated by a single world line, with a certain number of multipolar fields defined all along it).
It is customary to identify a \lq\lq global coordinate system" $x^\alpha=(ct, x^i)$ with the origin at the center of mass (COM) of the whole system and \lq\lq $N$ local coordinate systems" each attached with a single body. They are needed to split the general problem into two parts: the \lq\lq external problem" aimed at determining the motion of the COMs of the $N$ bodies, and the \lq\lq internal problem" aimed at determining the motion of each body around its COM. 
In the global coordinate system, the parametric equations of the $N$ COM world lines ${\mathcal L}_{A}$ ($A=1,\ldots, N$) have equations
$z_{A}^\alpha=z_{A}^\alpha(\tau_{A})$,  where $\tau_{A}$ is the proper time parametrization along each world line. 
Local coordinate systems, instead, can be, e.g., Fermi coordinates $X_{A}^\alpha=(cT_{A}, X_{A}^a)$ along the lines ${\mathcal L}_{A}$ (or any similar set of attached coordinates to these lines).
A mapping between the global and local set of coordinates is discussed in Ref. \cite{Damour:1990pi}.

If the spacetime region of interest never gets too close to any specific body, the gravitational field can be considered a perturbation of the flat spacetime metric $\eta_{\alpha\beta}$ referred to standard Cartesian coordinates, i.e.,
\beq
\label{backgroundmet}
g^{(0)}_{\alpha\beta}=\eta_{\alpha\beta}+h_{\alpha\beta}^{\rm M}\,,
\eeq
where $h_{\alpha\beta}^{\rm M}$ denotes the gravitational field of the Solar System bodies, i.e., the \lq\lq matter'' (M) field, and can be expressed either in the global coordinates $x^\mu$ or in any of the $N$ local coordinates $X^\mu_{A}$.
We will work to first order in $h$ throughout the paper.

\subsection{Perturbation by an incoming plane gravitational wave}

Let us consider a perturbation of this system induced by a plane gravitational wave emitted by a distant source.
The complete metric (background plus perturbation) is then given by 
\beq
g_{\alpha\beta}=\eta_{\alpha\beta}+h_{\alpha\beta}^{\rm M}+h_{\alpha\beta}^{\rm GW}\,,
\eeq
where both M and GW metrics can be treated as independent first order corrections to the Minkowski metric.
For the sake of simplicity we will take the metric of a monochromatic plane wave, which has the form 
\beq
h_{\alpha\beta}^{\rm GW}-\frac12\eta_{\alpha\beta}h^{\rm GW}{}^{\gamma}{}_{\gamma}={\rm Re}[A_{\alpha\beta}e^{ik_\alpha x^\alpha}]\,,
\eeq
where $A_{\alpha\beta}$ is a constant symmetric tensor, the polarization tensor, and $k$ is a constant null vector, the wave vector.
We will adopt the transverse-traceless (TT) gauge, so that $h_{\alpha\beta}^{\rm GW}$ is traceless (i.e., $h^{\rm GW}{}^{\alpha}{}_{\alpha}=0$) with nonvanishing components only on the plane orthogonal to the direction $k$ of propagation of the wave (i.e., $h_{\alpha\beta}^{\rm GW}k^\beta=0$), implying that $h_{0\beta}^{\rm GW}=0$.
The polarization tensor has only two independent components, corresponding to the two possible polarization states.

The gravitational fields associated with GWs are assumed weak enough to be considered in the linear approximation. Their effects on the deflection of light add linearly to those due to the Solar System bodies.

\subsection{Fiducial observers and adapted frames}

It is useful to introduce on the spacetime manifold an observer family $u$, which forms a congruence of timelike world lines characterized by the kinematical quantities acceleration $a(u)$, expansion $\theta(u)$, and vorticity $\omega(u)$ \cite{Jantzen:1992rg}, resulting from the splitting of its covariant derivative $\nabla u\equiv\nabla_\beta u_\alpha=u_{\alpha;\beta}$, i.e.,
\begin{eqnarray}
 a (u)_\alpha &=& u_{\alpha;\beta}u^\beta \,,\nonumber \\
 \theta (u)_{\alpha\beta} &=& P(u)^\gamma {}_\alpha
  P(u)^\delta {}_\beta\,u_{(\gamma;\delta)}\,,\nonumber \\
 \omega (u)_{\alpha\beta} &=&- P(u)^\gamma {}_\alpha
  P(u)^\delta {}_\beta\,u_{[\gamma;\delta]} \,,
\end{eqnarray}
where $P(u)^\alpha {}_\beta=\delta^\alpha{}_{\beta}+u^\alpha u_\beta$ projects orthogonally to $u$.
Let us assume as fiducial observers those at rest with respect to the global coordinates, i.e., the static observers, with associated $4$-velocity 
\beq
\label{obser}
u=u^\alpha\partial_\alpha=\left(1+\frac12 h_{00}\right)\partial_0\,,
\eeq
and kinematical fields with coordinate components
\begin{eqnarray}
\label{mkinq}
a(u)^i
&=&h_{0i,0}-\frac12 h_{00,i}\,,\nonumber\\
\theta(u)_{ij}
&=&\frac12 h_{ij,0}\,,\nonumber\\
\omega(u)_{ij}
&=&-h_{0[i,j]}\,,
\end{eqnarray}
so that for example $a(u)=a(u)^i\partial_i$, $\theta(u)=\theta(u)_{ij}\rmd x^i\otimes \rmd x^j$, etc.
An observer-adapted orthonormal spatial frame results in the following three vectors
\begin{eqnarray}
\label{fram}
e(u)_{\hat x}&=& h_{0x}\partial_0+\left(1-\frac12 h_{xx}\right)\partial_x\,,\\
e(u)_{\hat y}&=& h_{0y}\partial_0-h_{xy}\partial_x+\left(1-\frac12 h_{yy}\right)\partial_y\,, \nonumber\\
e(u)_{\hat z}&=& h_{0z}\partial_0-h_{xz}\partial_x-h_{yz}\partial_y+\left(1-\frac12 h_{zz}\right)\partial_z
\,.\nonumber
\end{eqnarray}

\subsection{Photon motion}

Every astrometric model should be able to reconstruct the trajectory of a light ray detected by an observer back to the source, i.e., to solve the inverse ray-tracing problem (see, e.g., Ref. \cite{Bini:2017apc} for a recent fully explicit application), up to the requested level of accuracy. In GREM the photon trajectory is parametrized by the coordinate time $t$, so that the null geodesic equations reduce to a set of second order ordinary differential equations for the spatial variables $x^a(t)$. These equations are solved by imposing mixed initial-boundary conditions by fixing the spatial coordinates of the photon at the time of emission and the unit tangent vector to the light trajectory in the infinite past, i.e., at infinite spatial distance from the origin of the global coordinate system \cite{Klioner:1991,Klioner:1992,Kopeikin:1999ev}.
Differently, RAMOD provides a set of (equivalent) first order ordinary differential equations for the coordinate components of the spatial light direction in the rest frame of a local static observer, i.e., its line of sight at each point of the light trajectory, as functions of a suitably defined non-affine parameter along the path \cite{deFelice:2004nf,deFelice:2006xm}. Further integration gives the coordinate position of the star. These equations are integrated by imposing boundary conditions at the time of observation in terms of the angular directions of the incoming light ray with respect to the spatial axes of a frame comoving with the satellite and the coordinate position of the satellite's trajectory (see, e.g., Ref. \cite{deFelice:2006jm} and references therein for additional details). 

Let $K$ be the tangent vector to the photon null geodesic world line, i.e.,
\beq
\label{geo_gen}
K^\alpha\nabla_\alpha K^\beta=0\,, \qquad
K^\alpha K_\alpha=0\,, 
\eeq
parametrized by the affine parameter $\lambda$ such that $K^\alpha={\rmd x^\alpha}/{\rmd \lambda}$.
We will use the following decomposition of the photon 4-momentum with respect to any given observer family $u$ \cite{deFelice:2004nf,deFelice:2006xm} 
\beq
\label{Kramod}
K=-(u\cdot K)u+\ell(u)
\equiv {\mathcal E}(K,u)u+\ell(u)\,,
\eeq
with $\ell(u)^\alpha=P(u)^\alpha{}_{\beta}K^\beta$ the observer-relative (spatial) momentum orthogonal to $u^\alpha$ and ${\mathcal E}(K,u)=-u\cdot K$ the observer-relative energy.

In place of $\lambda$ Refs. \cite{deFelice:2004nf,deFelice:2006xm} introduce another non-affine parameter $\sigma$ for the orbit, such that 
\beq
\bar K^\alpha=\frac{K^\alpha}{{\mathcal E}(K,u)}=\frac{\rmd x^\alpha}{\rmd \sigma}\,, 
\eeq
with
\beq
\label{barelldef}
\bar \ell(u) =\frac{\ell(u)}{{\mathcal E}(K,u)}=\bar K-u\,,
\eeq
a unit (spatial) vector representing the observer-relative direction of the momentum.
$\sigma$ is related to the affine parameter $\lambda$ by $\rmd \sigma={\mathcal E}(K,u)\rmd \lambda$, implying that Eq. (\ref{geo_gen}) becomes
\beq
\label{geo_sigma}
\bar K^\alpha\nabla_\alpha \bar K^\beta=-\left(\frac{\rmd}{\rmd\sigma}\ln{\mathcal E}(K,u)\right)\bar K^\beta\,.
\eeq
The observer-relative energy satisfies the equation
\beq
\frac{\rmd}{\rmd\sigma}\ln{\mathcal E}(K,u)
=-\bar\ell(u)^\alpha\bar\ell(u)^\beta \theta(u)_{\alpha\beta}-\bar\ell(u)^\alpha a(u)_\alpha\,,
\eeq
so that Eq. \eqref{geo_sigma} becomes

\begin{widetext}

\beq
\frac{\rmd \bar K^\alpha}{\rmd \sigma}+\Gamma^\alpha{}_{\mu\nu}\bar K^\mu\bar K^\nu
-\left[\bar\ell(u)^\mu\bar\ell(u)^\nu \theta(u)_{\mu\nu}+\bar\ell(u)^\mu a(u)_\mu\right]\bar K^\alpha=0\,,
\eeq
with $\Gamma^\alpha{}_{\mu\nu}=\frac12\eta^{\alpha\rho}\left(h_{\rho\nu,\mu}+h_{\rho\mu,\nu}-h_{\mu\nu,\rho}\right)$ evaluated along the photon path.
Equation \eqref{barelldef} then implies 
\begin{eqnarray}
\label{RME_gen}
\frac{\rmd \bar\ell(u)^\alpha}{\rmd \sigma}&+&
\Gamma^\alpha{}_{\mu\nu}\bar\ell(u)^\mu(\bar\ell(u)^\nu+u^\nu)+a(u)^\alpha-k(u)^\alpha{}_\sigma \bar\ell(u)^\sigma\nonumber\\
&-&\left[\bar\ell(u)^\mu\bar\ell(u)^\nu \theta(u)_{\mu\nu}+\bar\ell(u)^\mu a(u)_\mu\right](\bar\ell(u)^\alpha+u^\alpha)=0\,,
\end{eqnarray}
which is valid for any observer $u$.
Hereafter we will adopt the simplified notation $\bar\ell(u)^\alpha=\bar\ell^\alpha$, being understood that the direction of light propagation $\bar\ell$ is relative to the observer $u$. We will make explicit the dependence of $\bar\ell^\alpha$ and the metric components $h_{\alpha\beta}$ on the parameter $\sigma$ along the orbit, when convenient.

For the static observers with 4-velocity \eqref{obser} and kinematical fields \eqref{mkinq} the previous equations become
\begin{eqnarray}
\label{RME_static1}
\frac{\rmd\bar \ell^0}{\rmd \sigma}&=&\bar \ell^i\bar \ell^j h_{0i,j} + \bar \ell^i h_{0i,0}\,,\\
\label{RME_static2}
\frac{\rmd\bar \ell^k}{\rmd \sigma}&=&-\bar \ell^i\bar \ell^j \left(h_{ki,j} - \frac12h_{ij,k}\right)
-\bar \ell^i(h_{k0,i}+h_{ki,0}-h_{0i,k})
-h_{k0,0}+\frac12h_{00,k}\nonumber\\
&&
+\bar \ell^k\left[\frac12 \bar \ell^i\bar \ell^j h_{ij,0}+\bar \ell^i\left(h_{0i,0} - \frac12 h_{00,i}\right)\right]\,,
\end{eqnarray}

\end{widetext}
to be completed by 
\beq
\label{eqxRME}
\frac{\rmd x^0}{\rmd \sigma}=\bar \ell^0(\sigma)+1+\frac12 h_{00}(\sigma)\,,\qquad
\frac{\rmd x^a}{\rmd \sigma}=\bar \ell^a(\sigma)\,.
\eeq
These equations, valid through $O(h)$, have been derived in this form in Ref. \cite{deFelice:2006xm}.
Unfortunately, the equation for $\bar \ell^0$ is incorrect there (see Appendix B).

To first order $O(h)$ the solution of Eqs. \eqref{RME_static1}--\eqref{eqxRME} can be written as
\beq
\label{l0vsla}
\bar \ell^0=h_{0i}\bar \ell^i\,,\qquad
\bar\ell^a=\bar\ell^a_{\not\,0}+\bar\ell^a_{h}\,,\qquad
x^\alpha=x^\alpha_{\not\,0}+x^\alpha_{h}\,,
\eeq
where $\bar\ell^a_{\not\,0}$ denotes the unperturbed local photon direction and the parameter $\sigma$ on the light ray trajectory is fixed so that it is $\sigma=0$ at the event of observation (i.e., at the satellite position) with coordinates $x^\alpha_{\rm obs}=(x^0_{\rm obs},x^i_{\rm obs})$. 
The null condition to $O(h)$ then implies 
\beq
\label{nullcond}
2\bar\ell_{\not\,0}\cdot\bar\ell_{h}+h_{\bar\ell_{\not\,0}\bar\ell_{\not\,0}}=0\,, \qquad
h_{\bar\ell_{\not\,0}\bar\ell_{\not\,0}}=h_{ab}\bar\ell^a_{\not\,0}\bar\ell^b_{\not\,0}\,.
\eeq
Therefore, the unperturbed orbit (which is a straight line) can be written as
\beq
x^0_{\not\,0}=x^0_{\rm obs}+\sigma\,,\qquad
x^a_{\not\,0}=x^a_{\rm obs}+\bar\ell^a_{\not\,0}\sigma\,,
\eeq
with $x^\alpha_{h}(0)=0$.
The \lq\lq actual'' (locally spatial) photon direction evaluated at the observation point $\bar\ell^a_{\rm obs}=\bar\ell^a_{\not\,0}+\bar\ell^a_{h}(0)$ is considered fully known, being related to direct observations and to the selected attitude of the observer's frame.
Therefore, the solution of Eqs. \eqref{RME_static1}--\eqref{eqxRME} can be written as
\begin{eqnarray}
\label{x0sol}
x^0(\sigma)&=&x^0_{\rm obs}+\sigma+x^0_{h}(\sigma)\,,\\
\label{xasol}
x^a(\sigma)&=&x^a_{\rm obs}+\bar\ell^a_{\not\,0}\sigma+x^a_{h}(\sigma)\nonumber\\
&=&x^a_{\rm obs}+\bar\ell^a_{\rm obs}\sigma+(x^a_{h}(\sigma)-\bar\ell^a_{h}(0)\sigma)\,,\\
\label{lasol}
\bar\ell^a(\sigma)&=&\bar\ell^a_{\rm obs}+(\bar\ell^a_{h}(\sigma)-\bar\ell^a_{h}(0))\,.
\end{eqnarray}
Eq. \eqref{x0sol} can be used in turn to switch to the coordinate time parametrization of GREM, whereas evaluating Eq. \eqref{xasol} at the spatial position $x^a_*=x^a(\sigma_*)$ of the star gives the components $\bar\ell^a_{\rm obs}$ at the observer's position in terms of the star coordinates (after eliminating the parameter $\sigma_*$ through the normalization condition \eqref{nullcond}).

\subsection{Astrometric observables}

\subsubsection{Satellite adapted frame}

The (timelike) satellite world line has 4-velocity $U$,
\beq
\label{sat_4vel}
U=\gamma(U,u)[u+\nu(U,u)]
=\Gamma [\partial_0+v^a\partial_a]\,,
\eeq
where $\Gamma$ is a normalization factor and
\begin{eqnarray}
\nu(U,u)&=&||\nu(U,u)||\hat\nu(U,u)
=\nu(U,u)^{\hat a}e(u)_{\hat a}\,,\nonumber\\
\gamma(U,u)&=&(1-||\nu(U,u)||^2)^{-1/2}\,,
\end{eqnarray}
so that $U\cdot U=-1$ and $v^a$ depend on $t$ only. 

An adapted frame to this world line can be obtained by boosting the orthonormal threading frame $\{u,e(u)_{\hat a}\}$ along $U$, i.e.,
\begin{eqnarray}
\label{satellite_frame2}
E_{\hat 0}&=&U\,,\nonumber\\
E_{\hat a}&=&e(u)_{\hat a}+\frac{(U\cdot e(u)_{\hat a})}{\gamma(U,u)+1}(U+u)\,,
\end{eqnarray}
with $E_{\hat a}\cdot U=0$, are the axes $e(u)_{\hat a}$ of the observer $u$ as seen by the observer $U$.
The normalization factor $\Gamma$ in Eq. \eqref{sat_4vel} is given by 
\beq
\Gamma 
= \Gamma_0\left[1+\Gamma_0^2\left(\frac12h_{00}+h_{0v}+\frac12h_{vv}\right)\right]\,,
\eeq
with $\Gamma_0=(1-v^2)^{-1/2}$, where the notation $h_{0v}=h_{ta}v^a$, $h_{vv}=h_{ab}v^av^b$ and $v^2=\delta_{ab}v^av^b$ has been used.
In terms of the coordinate components of the spatial velocity the frame components $\nu(U,u)^{\hat a}$ and the associated Lorentz factor $\gamma(U,u)$ are 
\begin{eqnarray}
\label{satvelframe}
\nu(U,u)^{\hat x}&=&v^x\left[1+\frac12\left(h_{00}+h_{xx}\right)+h_{0v}\right]+h_{xy}v^y\nonumber\\
&&
+h_{xz}v^z\,,\nonumber\\
\nu(U,u)^{\hat y}&=&v^y\left[1+\frac12\left(h_{00}+h_{yy}\right)+h_{0v}\right]+h_{yz}v^z\,,\nonumber\\
\nu(U,u)^{\hat z}&=&v^z\left[1+\frac12\left(h_{00}+h_{zz}\right)+h_{0v}\right]\,,\nonumber\\
\gamma(U,u)&=& \Gamma_0\left\{1+\Gamma_0^2\left[\left(\frac12h_{00}+h_{0v}\right)v^2+\frac12h_{vv}\right]\right\}
\,,\nonumber\\
\end{eqnarray}
to first order in $h$.

The satellite attitude frame is specified by a suitable spatial rotation of the adapted triad (\ref{satellite_frame2}) 
\beq
\label{Fframe}
F_{\hat a}=E_{\hat b}{\mathcal R}^{\hat b}{}_{\hat a}\,, 
\eeq
where the rotation matrix can be equivalently parametrized by either three Euler angles, or the Cayley-Klein parameters or even the modified Rodrigues parameters (see, e.g., Eq. (53) in Ref. \cite{Crosta:2017eco} for Gaia).
We will refer to them as attitude parameters $a_i$.

\subsubsection{Main observables}

Following the notation of Ref. \cite{deFelice:2006xm}, let 
\begin{eqnarray}
\label{cospsidef}
\cos\psi_{({\hat a},{K})}&=&\frac{K\cdot F_{\hat a}}{(-U\cdot K)}\bigg\vert_{\rm obs}\\
&=&\frac{\bar\ell(u)\cdot F_{\hat a}}{\gamma(U,u)(1-\nu(U,u)\cdot\bar\ell(u))}\bigg\vert_{\rm obs}\,,\nonumber
\end{eqnarray}
be the $\hat a$-th direction cosine measured by the observer on the satellite.
Each observation can be translated in the measurement of two coordinates, $\cos\phi$ on the focal plane (along-scan), and $\sin\zeta$ orthogonal to it (across-scan), which are related to the direction cosines \eqref{cospsidef} by \cite{Crosta:2017eco}
\beq
\cos\phi=\frac{\cos\psi_{({\hat 1},{K})}}{|\sin\psi_{({\hat 3},{K})}|}\,, \qquad
\sin\zeta=\cos\psi_{(\hat 3,K)}\,.
\eeq
Repeated observations of the same objects from different satellite orientations and at different times allow to estimate their astrometric parameters, i.e., angular positions, parallaxes, and proper motions, as shown in Ref. \cite{Crosta:2017eco}. From a computational point of view, these observations produce a large number of equations, which is much larger than the number of unknowns. 
Furthermore, the latter enter the observation equations in a highly nonlinear way in general, so that a nonlinear optimization algorithm should be needed. However, assuming that the initial values of all unknown parameters are close enough to the true ones, one can linearize the system of equations with respect to the unknowns around a known set of reference values. The solution through a least-squares method eventually provides the catalog and associated uncertainties (we refer to Ref. \cite{Crosta:2017eco} and references therein for a detailed account of the Gaia data analysis).  

We will write the solution for the direction cosines as the sum of two contributions, i.e., 
\beq
\cos\psi_{({\hat a},{K})}=[\cos\psi_{({\hat a},{K})}]^{(0)}+[\cos\psi_{({\hat a},{K})}]^{\rm GW}\,,
\eeq
where $[\cos\psi_{({\hat a},{K})}]^{(0)}$ is the part of $\cos\psi_{({\hat a},{K})}$ due to the \lq\lq background'' field \eqref{backgroundmet}, i.e., to the gravitational field of the Solar System, and $[\cos\psi_{({\hat a},{K})}]^{\rm GW}$ is the correction due to the gravitational wave.

\section{The gravitational field of the Solar System}

The Solar System is assumed to be isolated and described by the following metric in the Barycentric Celestial Reference System (BCRS) as recommended by the IAU resolution B1.3 (2000) \cite{Soffel:2003cr} 
\begin{eqnarray}
\label{BCRSmet}
g^{(0)}_{00}(t,x^i)&=& -1+ 2 \epsilon^2V- 2 \epsilon^4V^2+O(6)\,,\nonumber\\
g^{(0)}_{0i}(t,x^i)&=& -4 \epsilon^3 V^i+O(5)\,, \nonumber\\
g^{(0)}_{ij}(t,x^i)&=& \delta_{ij}[1+2\epsilon^2V]+O(4)\,,
\end{eqnarray}
where $\epsilon=1/c$ and $O(n)=O(\epsilon^n)$, the functions $V=V(t,x^i)$ and $V_i=V_i(t,x^i)$ denoting the potentials associated with the gravitational field.
The BCRS coordinates are harmonic, since the metric \eqref{BCRSmet} satisfies the gauge conditions $\partial_\beta (\sqrt{-g^{(0)}} g^{(0)\,\alpha\beta})=0=g^{(0)\,\mu\nu}\Gamma^{(0)\,\alpha}{}_{\mu\nu}$, that is
\beq
\partial_t V+ \partial_i V^i=0(4)\,.
\eeq
The gravitational potentials $V$ and $V^i$ are given by 
\beq
V(t,x^i)=\sum_{A=1}^N V_A(t,x^i)\,,\qquad
V^i(t,x^i)=\sum_{A=1}^N V^i_A(t,x^i)\,,
\eeq
where the individual contribution of the $A$-th body is computed in the local coordinate system $(cT_A,X^i_A)$ attached with it in terms of two families of intrinsic multipole moments, mass $M_L$ and spin $S_L$ moments, defined through the associated energy-momentum tensor \cite{Damour:1990pi}. 

In order to meet the microarcsecond level of accuracy it is enough to keep terms in the metric \eqref{BCRSmet} up to the order $O(3)$ included.
Furthermore, the sources can be assumed to move with a constant velocity relative to the global reference system.
They can also be considered as nonrotating and endowed with a quadrupolar structure, with a time-independent mass quadrupole moment.
It is enough to take into account the contribution of giant planets only, which can be modeled as (flattened) homogeneous axisymmetric ellipsoids of revolution, whose principal axes are aligned with the spatial axes of the global coordinate system. 
Such approximations are extensively discussed in Refs. \cite{Klioner:1991,Klioner:1992,Kopeikin:1999ev,Klioner:2003,Zschocke:2010qn,Zschocke:2015mca}.

The gravitational potentials of the $A$-th source are then given by
\beq
\label{potdef}
V_{A}(t,x^i)=h_{A}(t,x^i)\,, \qquad
V^i_{A}(t,x^i)=h_{A}(t,x^i)\tilde v^i_{A}\,,
\eeq
with \cite{Klioner:2003}
\beq
\label{hquaddef}
h_{A}= \frac{GM_{A}}{r_{A}} \left[ 1- J_{2A}\left(
\frac{R^{\rm eq}_{A}}{r_{A}} \right)^2  P_2\left(\frac{z-z_{A}}{r_{A}}\right) \right]\,,
\eeq
where $r^i_{A}= x^i - x^i_{A}(t)$ are the coordinates of the $A$-th body with the origin fixed at the center of mass of the whole system, so that 
\beq
r_{A}(t,x^i)=\left[(x-x_A(t))^2+(y-y_A(t))^2+(z-z_A(t))^2\right]^{1/2}\,, 
\eeq
with $x^i_{A}(t)=x^i_{A}(t_{A,0})+\tilde v^i_{A}(t-t_{A,0})$, $t_{A,0}$ denoting a given reference time.
$P_2(x)=\frac12(3x^2-1)$ is the Legendre polynomial of degree $n=2$, $M_{A}$ the mass of the $A$-th body, $R^{\rm eq}_{A}$ its equatorial radius, and the coefficient $J_{2A}$ the dimensionless mass quadrupole parameter.
In the following we will remove the subscript $A$ for simplicity and we will retain terms up to the order $O(3)$ in the expansion of perturbation quantities.

The solution of the photon equations of motion is then given by \cite{Crosta:2017eco}
\begin{eqnarray}
\label{solbarlfin}
\bar\ell^a(\sigma)&=& \bar\ell^a_{\rm obs}
+\epsilon^2\left\{2[1-2\epsilon\vdotl]H^a(\sigma)\right.\nonumber\\
&&\left.
-(3\bar\ell^a_{\not\,0}-4\epsilon\tilde v^a)(h(\sigma)-h_{\rm obs})\right\}
+2\epsilon^3\bar\ell^a_{\not\,0}H^t(\sigma)\nonumber\\
&=&\bar\ell^a_{\not\,0}+\bar\ell_h^{{\rm M}\,a}
\,,
\end{eqnarray}
with $\vdotl=\delta_{ab}\tilde v^a\bar\ell^b_{\not\,0}$, and
\begin{eqnarray}
\label{solxfin}
x^0(\sigma)&=&x^0_{\rm obs}+\sigma+\epsilon^2[1-4\epsilon\vdotl]H(\sigma)\,,\nonumber\\
x^a(\sigma)&=&x^a_{\rm obs}+\bar\ell^a_{\rm obs}\sigma
+\epsilon^2\left\{2[1-2\epsilon\vdotl]{\mathcal H}^a(\sigma)\right.\nonumber\\
&&\left.
-(3\bar\ell^a_{\not\,0}-4\epsilon\tilde v^a)(H(\sigma)-h_{\rm obs}\sigma)\right\}
+2\epsilon^3\bar\ell^a_{\not\,0}{\mathcal H}^t(\sigma)
\,,\nonumber\\
\end{eqnarray}
where we have introduced the quantities 
\begin{eqnarray}
\label{Hdefs}
H(\sigma)&=&\int_0^{\sigma}h(\sigma)\rmd\sigma\,,\qquad
H^a(\sigma)=\int_0^{\sigma}[\partial_ah](\sigma)\rmd\sigma\,,\nonumber\\
{\mathcal H}^a(\sigma)&=&\int_0^{\sigma}H^a(\sigma)\rmd\sigma\,,\\
H^t(\sigma)&=&\int_0^{\sigma}[\partial_th](\sigma)\rmd\sigma\,,\qquad
{\mathcal H}^t(\sigma)=\int_0^{\sigma}H^t(\sigma)\rmd\sigma\,.\nonumber
\end{eqnarray}
Finally, the normalization factor ${\mathcal E}$ turns out to be
\beq
\label{calEsol}
{\mathcal E}(\sigma)=1+\epsilon^2(h(\sigma)-h_{\rm obs})-2\epsilon^3H^t(\sigma)\,, 
\eeq
where the unperturbed value has been set equal to unity without any loss of generality.

A solution to the photon equations of motion \eqref{RME_static1}--\eqref{eqxRME} in the context of Gaia was already obtained in Ref. \cite{Crosta:2015jia} for uniformly moving quadrupolar bodies, by using a different formulation.
However, it is incorrect, as shown in Appendix B below.
The correct solution in the case of pointlike sources moving with constant velocities was recently presented in Ref. \cite{Crosta:2017eco} in the same form as Eqs. \eqref{solbarlfin}--\eqref{calEsol}, with the functions \eqref{Hdefs} given in Appendix B there.
We will provide in Appendix A below the correct solution for extended bodies also endowed with a mass quadrupole moment, so improving the results of Ref. \cite{Crosta:2017eco} and fully correcting those of Ref. \cite{Crosta:2015jia}.

\subsection{Astrometric observables}

The spatial triad (\ref{satellite_frame2}) adapted to the satellite world line $U$ becomes in this case 
\begin{eqnarray}
\label{satellite_frame1PN}
E_{\hat a}&=& \left\{\epsilon v^a +\epsilon^3 \left[v^a\left(\frac{v^2}{2}+3h\right)  -4h\tilde v^a\right]\right\}\partial_0\nonumber\\
&& 
+\left(1-\epsilon^2 h\right)\partial_a
+\frac{\epsilon^2}{2}v^av^b\partial_b+ O(4)\,.
\end{eqnarray}
The direction cosines \eqref{cospsidef} are thus given by 
\beq
[\cos\psi_{({\hat a},{K})}]^{(0)}=[\cos\psi_{({\hat a},{K})}]^{\rm flat}+[\cos\psi_{({\hat a},{K})}]^{\rm M}\,,
\eeq
where

\begin{widetext}

\begin{eqnarray}
[\cos\psi_{({\hat a},{K})}]^{\rm flat}&=&
\boldsymbol{C}_{\hat a}\cdot \boldsymbol{\bar\ell}_{\not\,0}
+\epsilon[(\boldsymbol{v}\cdot \boldsymbol{\bar\ell}_{\not\,0})(\boldsymbol{C}_{\hat a}\cdot \boldsymbol{\bar\ell}_{\not\,0})-\boldsymbol{C}_{\hat a}\cdot \boldsymbol{v}]\nonumber\\
&&
+\epsilon^2[1+\epsilon(\boldsymbol{v}\cdot \boldsymbol{\bar\ell}_{\not\,0})]\left\{
-\frac12(\boldsymbol{v}\cdot \boldsymbol{\bar\ell}_{\not\,0})(\boldsymbol{C}_{\hat a}\cdot \boldsymbol{v})+(\boldsymbol{C}_{\hat a}\cdot \boldsymbol{\bar\ell}_{\not\,0})\left[(\boldsymbol{v}\cdot \boldsymbol{\bar\ell}_{\not\,0})^2-\frac{v^2}2\right]
\right\}+O(4)\,,
\end{eqnarray}
is the flat spacetime value, whereas 
\begin{eqnarray}
[\cos\psi_{({\hat a},{K})}]^{\rm M}&=&
\boldsymbol{C}_{\hat a}\cdot \boldsymbol{\bar\ell}_h^{\rm M}
+\epsilon[(\boldsymbol{C}_{\hat a}\cdot \boldsymbol{\bar\ell}_{\not\,0})(\boldsymbol{v}\cdot \boldsymbol{\bar\ell}_h^{\rm M})+(\boldsymbol{v}\cdot \boldsymbol{\bar\ell}_{\not\,0})(\boldsymbol{C}_{\hat a}\cdot \boldsymbol{\bar\ell}_h^{\rm M})]\nonumber\\
&&
+\epsilon^2h[(\boldsymbol{C}_{\hat a}\cdot \boldsymbol{\bar\ell}_{\not\,0})(1+4\epsilon(\boldsymbol{v}\cdot \boldsymbol{\bar\ell}_{\not\,0}))-2\epsilon(\boldsymbol{C}_{\hat a}\cdot \boldsymbol{v})]+O(4)\,,
\end{eqnarray}

\end{widetext}
is the first order correction due to the \lq\lq matter'' field, 
which has to be evaluated at the position of the satellite, i.e., for $\sigma=0$.
Here the notation $(\boldsymbol{A}\cdot \boldsymbol{B})=\delta_{ab}A^aB^b$ has been used for the scalar product between three-dimensional vectors. 
The coefficients $C^{b}_{\hat a}={\mathcal R}^{b}{}_{\hat a}$ are all functions of the attitude parameters only.

Therefore, the $\hat a$-th direction cosine turns out to be a function of the spatial position $x^a_*$ of the star (or, equivalently, its astrometric parameters) and the satellite's attitude represented by the parameters $a_i$, i.e., $[\cos\psi_{({\hat a},{K})}]^{(0)}=f_{\hat a}(x^i_*,a_i)$.
The variation of this equation with respect to the parameters is easily computed (see Section IV of Ref. \cite{Crosta:2017eco}), leading to a linearized set of equations around a known solution at the time of observation, which is then solved by using the least-square method, as stated above.

\section{Astrometric effects induced by a gravitational plane wave}

Let us consider the perturbation due to a monochromatic gravitational plane wave with frequency $\omega$ travelling along an arbitrary direction (with wave vector $k=\omega\partial_0+k^a\partial_a$). In the TT-gauge the associated metric has nonvanishing components $h_{ab}^{\rm GW}=O(4)$ with (see, e.g., Ref. \cite{Klioner:2017asb})
\begin{eqnarray}
\label{GWmet}
h_{ab}^{\rm GW}\rmd x^a\rmd x^b&=& (\alpha h_+-\beta h_\times)\rmd x^2 +(\gamma h_++\beta h_\times)\rmd y^2\nonumber\\
&& 
-h_+\rmd z^2\nonumber\\
&& 
+2\left(-\frac{k_y}{k_x}\gamma h_++\frac{k_z}{k_x}\delta h_\times\right)\rmd x\rmd y\nonumber\\
&&  
+2\left(\frac{k_z}{k_x}h_++\frac{k_y}{k_x} h_\times\right)\rmd x\rmd z\nonumber\\
&&  
-2h_\times \rmd y \rmd z\,,
\end{eqnarray}
provided that $k_x\not=0$, 
with coefficients
\begin{eqnarray}
\alpha&=&\frac{k_y^2-k_z^2}{k_x^2+k_y^2}\,,\qquad
\beta=\frac{2k_yk_z}{k_x^2+k_y^2}\,,\nonumber\\
\gamma&=&1-\alpha\,,\qquad
\delta=\frac{k_x^2-k_y^2}{k_x^2+k_y^2}\,,
\end{eqnarray}
and polarization functions $h_+=h_+^s-h_+^c$ and $h_\times=h_\times^c-h_\times^s$, with
\begin{eqnarray}
h_+^c&=&A_+^c \cos W\,,\qquad 
h_+^s=A_+^s \sin W\,,\nonumber\\ 
h_\times^c&=&A_\times^c \cos W \,, \qquad
h_\times^s=A_\times^s \sin W
\,,
\end{eqnarray}
and $W=k\cdot x=k_\alpha x^\alpha$.
The two GW polarizations are thus equivalently parametrized by four strain amplitudes instead of two amplitudes and two phases. 
Note that there exist many equivalent forms of the metric \eqref{GWmet} depending on the chosen parametrization of the direction of propagation of the wave (see, e.g., Ref. \cite{Pyne:1995iy}, where spherical-like coordinates are used instead).

The parametric equations of null geodesic orbits with tangent vector $K$ (see Eq. \eqref{geo_gen}) in the metric $\eta_{\alpha\beta}+h_{\alpha\beta}^{\rm GW}$ are given by
\begin{eqnarray}
\label{geogw}
x^0(\lambda)&=&x^0_0+(E+B^t)\lambda+C^t_c(\cos W(\lambda)-\cos W_0)\nonumber\\
&&
+C^t_s(\sin W(\lambda)-\sin W_0)
\,, \nonumber\\
x^a(\lambda)&=&x^a_0+(K^a_0+B^a)\lambda+C^a_c(\cos W(\lambda)-\cos W_0)\nonumber\\
&&
+C^a_s(\sin W(\lambda)-\sin W_0)\,, 
\end{eqnarray}
where the unperturbed 4-momentum is denoted by $K_0=E\partial_0+K_0^a\partial_a$, $\lambda$ is an affine parameter and $x^\alpha_0=x^\alpha(\lambda=0)$, so that 
\beq
W(\lambda)=(K_0\cdot k)\lambda+W_0\,, \qquad
W_0=W(0)=k\cdot x_0\,,
\eeq
provided that $(K_0\cdot k)\not=0$, and

\begin{widetext}

\begin{eqnarray}
C^t_c&=&-\frac{\omega}{2k_x(K_0\cdot k)^2}\{
k_x[(\alpha A_+^s+\beta A_\times^s)K_{0x}^2+(-\beta A_\times^s+\gamma A_+^s)K_{0y}^2+2K_{0y}K_{0z}A_\times^s-K_{0z}^2A_+^s]\nonumber\\
&&
-2K_{0x}[(k_y\gamma A_+^s+k_z\delta A_\times^s)K_{0y}+(-A_+^sk_z+A_\times^sk_y)K_{0z}]
\}
\,,\nonumber\\
C^x_c&=&\frac{k_x}{\omega}C^t_c
-\frac{(K_{0z}k_y+K_{0y}k_z\delta-K_{0x}k_x\beta)A_\times^s-A_+^s(K_{0x}k_x\alpha-\gamma k_yK_{0y}+K_{0z}k_z)}{k_x(K_0\cdot k)}
\,,\nonumber\\
C^y_c&=&\frac{k_y}{\omega}C^t_c
+\frac{[(K_{0z}-K_{0y}\beta)k_x-K_{0x}k_z\delta]A_\times^s-A_+^s\gamma(K_{0x}k_y-K_{0y}k_x)}{k_x(K_0\cdot k)}
\,,\nonumber\\
C^z_c&=&\frac{k_z}{\omega}C^t_c
+\frac{(-K_{0x}k_y+K_{0y}k_x)A_\times^s+A_+^s(K_{0x}k_z-k_xK_{0z})}{k_x(K_0\cdot k)}
\,,
\end{eqnarray}

\end{widetext}
with $C^\alpha_s=C^\alpha_c(A_+^s\to A_+^c,A_\times^s\to A_\times^c)$, whereas $B^\alpha$ are $O(h)$ arbitrary constants such that $K_0\cdot B=0$.
The tangent vector $K$ to null geodesics is thus given by
\begin{eqnarray}
K&=&\left[1+(K_0\cdot k)(C^t_s \cos W-C^t_c\sin W)+B^t\right]\partial_0\nonumber\\
&&
+\left[K_0^a+(K_0\cdot k)(C^a_s \cos W-C^a_c\sin W)+B^a\right]\partial_a
\,,\nonumber\\
\end{eqnarray}
where we have set the unperturbed photon energy $E=1$ as before.

Let us introduce also in this case the decomposition \eqref{Kramod} of the photon 4-momentum with respect to the static observers $u=\partial_0$. 
We find ${\mathcal E}=1+{\mathcal E}^{\rm GW}$ for the normalization factor, with
\beq
{\mathcal E}^{\rm GW}=(K_0\cdot k)(C^t_s \cos W(\lambda)-C^t_c\sin W(\lambda))+B^t\,,
\eeq
so that the non-affine parameter $\sigma$ parametrizing the photon trajectory turns out to be
\begin{eqnarray}
\label{sigmasolgw}
\sigma&=&(1+B^t)\lambda+C^t_c(\cos W(\lambda)-\cos W_0)\nonumber\\
&&
+C^t_s(\sin W(\lambda)-\sin W_0)\,,
\end{eqnarray}
which can be easily inverted to yield $\lambda$ as a function of $\sigma$.
The arbitrary constant $B^t$ can then be chosen so that ${\mathcal E}=1$ for $\sigma=0$, i.e., $B^t=-(K_0\cdot k)(C^t_s \cos W_0-C^t_c\sin W_0)$, leading to
\begin{eqnarray}
{\mathcal E}^{\rm GW}(\sigma)&=&-(K_0\cdot k)[C^t_s(\cos W(\sigma)-\cos W_0)\nonumber\\
&&
+C^t_c (\sin W(\sigma)-\sin W_0)]\,.
\end{eqnarray}
Furthermore,
\begin{eqnarray}
\label{barellgw}
\bar \ell(\sigma)
&=&\left[K_0^a-(C^a_s-C^t_sK_0^a)\cos W(\sigma)\right.\nonumber\\
&&\left.
+(C^a_c-C^t_c K_0^a)\sin W(\sigma)\right]\partial_a\nonumber\\
&\equiv&\left(K_0^a+\bar\ell_h^{{\rm GW}\,a}(\sigma)\right)\partial_a
\,,
\end{eqnarray}
and
\begin{eqnarray}
\label{geogwsigma}
x^0(\sigma)&=&x^0_0+\sigma
\,, \\
x^a(\sigma)&=&x^a_0+K_0^a\sigma\nonumber\\
&&
+(C^a_c-C^t_cK_0^a)(\cos W(\sigma)-\cos W_0)\nonumber\\
&&
+(C^a_s-C^t_s K_0^a)(\sin W(\sigma)-\sin W_0) 
\,,\nonumber
\end{eqnarray}
where $x^\alpha_0=x^\alpha_{\rm obs}$, $K_0^a=\bar\ell^a_{\not\,0}$ and we have set $B^a=B^tK_0^a$.

\subsection{Astrometric observables}

The spatial triad (\ref{satellite_frame2}) adapted to the satellite world line $U$ in this case reads 
\begin{eqnarray}
\label{satellite_frame1PNGW}
E_{\hat a}=e(u)_{\hat a}+v^a\left(1+\frac{v^2}{2}\right)\partial_0
+\frac12v^av^b\partial_b\,, 
\end{eqnarray}
where 
\begin{eqnarray}
\label{frameGW}
e(u)_{\hat x}&=& \left(1-\frac12 h^{\rm GW}_{xx}\right)\partial_x\,,\\
e(u)_{\hat y}&=& -h^{\rm GW}_{xy}\partial_x+\left(1-\frac12 h^{\rm GW}_{yy}\right)\partial_y\,, \nonumber\\
e(u)_{\hat z}&=& -h^{\rm GW}_{xz}\partial_x-h^{\rm GW}_{yz}\partial_y+\left(1-\frac12 h^{\rm GW}_{zz}\right)\partial_z
\,,\nonumber
\end{eqnarray}
as from Eq. \eqref{fram} with $h_{0a}=0$.
The leading-order correction to the direction cosines then turns out to be
\beq
[\cos\psi_{({\hat a},{K})}]^{\rm GW}=\boldsymbol{C}_{\hat a}\cdot \boldsymbol{\hat\nu}_{h\,{\rm obs}}^{\rm GW}\,,
\eeq
where $(\boldsymbol{C}_{\hat a}\cdot \boldsymbol{\hat\nu}_{h\,{\rm obs}}^{\rm GW})=\delta_{ab}C^{a}_{\hat a}\hat\nu_{h\,{\rm obs}}^{{\rm GW}\,b}$ and
\begin{eqnarray}
\hat\nu_h^{{\rm GW}\,x}&=&\bar\ell_h^{{\rm GW}\,x}+\frac12h_{xx}^{\rm GW}\bar\ell_{\not\,0}^{x}+h_{xy}^{\rm GW}\bar\ell_{\not\,0}^{y}+h_{xz}^{\rm GW}\bar\ell_{\not\,0}^{z}
\,,\nonumber\\
\hat\nu_h^{{\rm GW}\,y}&=&\bar\ell_h^{{\rm GW}\,y}+\frac12h_{yy}^{\rm GW}\bar\ell_{\not\,0}^{y}+h_{yz}^{\rm GW}\bar\ell_{\not\,0}^{z}
\,,\nonumber\\
\hat\nu_h^{{\rm GW}\,z}&=&\bar\ell_h^{{\rm GW}\,z}+\frac12h_{zz}^{\rm GW}\bar\ell_{\not\,0}^{z}\,,
\end{eqnarray}
which have to be evaluated at the position of the satellite, i.e., for $\sigma=0$, where
\beq
\bar \ell_{h\,{\rm obs}}^{{\rm GW}\,a}=-(C^a_s-C^t_s\bar\ell_{\not\,0}^a)\cos W_0+(C^a_c-C^t_c \bar\ell_{\not\,0}^a)\sin W_0\,.
\eeq
Including terms which are linear in the satellite velocity we finally get
\begin{eqnarray}
[\cos\psi_{({\hat a},{K})}]^{\rm GW}&=&(\boldsymbol{C}_{\hat a}\cdot \boldsymbol{\hat\nu}_{h\,{\rm obs}}^{\rm GW})[1+\epsilon(\boldsymbol{v}\cdot \boldsymbol{\bar\ell}_{\not\,0})]\nonumber\\
&&
+\epsilon\left[
(\boldsymbol{C}_{\hat a}\cdot \boldsymbol{\bar\ell}_{\not\,0})\left(h^{\rm GW}_{\bar\ell_{\not\,0}v}+\left(\boldsymbol{v}\cdot \boldsymbol{\bar\ell}_{h\,{\rm obs}}^{\rm GW}\right)\right)\right.\nonumber\\
&&\left.
-(\boldsymbol{C}_{\hat a}\cdot \boldsymbol{\nu}_{h\,{\rm obs}}^{\rm GW})
\right]\,,
\end{eqnarray}
where $h^{\rm GW}_{\bar\ell_{\not\,0}v}=h^{\rm GW}_{ab}\bar\ell^a_{\not\,0}v^b$ and 
\begin{eqnarray}
\nu_h^{\hat x\,GW}&=&\frac12h_{xx}^{\rm GW}v^x+h_{xy}^{\rm GW}v^y+h_{xz}^{\rm GW}v^z
\,,\nonumber\\
\nu_h^{\hat y\,GW}&=&\frac12h_{yy}^{\rm GW}v^y+h_{yz}^{\rm GW}v^z
\,,\nonumber\\
\nu_h^{\hat z\,GW}&=&\frac12h_{zz}^{\rm GW}v^z\,,
\end{eqnarray}
are the GW-dependent part of the frame components \eqref{satvelframe} of the satellite's spatial velocity. 

Therefore, the $\hat a$-th direction cosine depends on seven further parameters: the four strain parameters $h_+^s,h_+^c,h_\times^s,h_\times^c$ and three parameters $k^a$ describing the direction of the gravitational wave (or equivalently two such parameters and the frequency $\omega$).
Unfortunately, in this case one does not know any initial value for any of these parameters, so that the least square method cannot be applied. 
A GW detection algorithm has been proposed in Ref. \cite{Klioner:2017asb}, using the technique of vector spherical harmonics \cite{Mignard:2012xm} and the HEALPix sky pixelization scheme \cite{Gorski:2004by}.

Let us conclude this section by comparing our results with those of Book and Flanagan \cite{Book:2010pf}, who first computed the change in the photon direction towards a distant astrometric source due to a plane gravitational wave. 
They considered an observer at rest, with an adapted frame parallel transported along his world line, so that our results cannot be directly related to those of Ref. \cite{Book:2010pf}. 
The photon 4-momentum is given there (Eq. (32)) by
\beq
\label{Kbf}
K=\omega_{\rm obs}\left(u-n^{\hat a}E_{\hat a}^{\rm(par)}\right)\,,
\eeq  
where $\omega_{\rm obs}$ denotes the observed photon frequency (Eq. (27)) and $n^{\hat a}=n^a+\delta n^{\hat a}$, $n$ denoting the unperturbed direction and $\delta n$ the $O(h)$ correction (Eq. (39)), and $\{u,E_{\hat a}^{\rm(par)}\}$ is a parallel transported frame along $u$ (Eqs. (30)--(31)), with
\beq
E_{\hat a}^{\rm(par)}=\left(\delta_a^b-\frac12h^{\rm GW}_{ab}\right)\partial_b\,,
\eeq
which is suitably rotated with respect to the spatial frame \eqref{satellite_frame1PNGW} (where one should also set $v^a=0$).
Direct comparison with the decomposition \eqref{Kramod} gives $\omega_{\rm obs}={\mathcal E}$ and 
\beq
\bar\ell^a=-n^a-\delta n^a+\frac12h^{\rm GW}_{ab}n^b\,,
\eeq
implying that $-n^a=K_0^a=\bar\ell^a_{\not\,0}$ and the $O(h)$ corrections to the coordinate components of the photon direction are related by
\beq
\bar\ell_h^{{\rm GW}\,a}=-\delta n^a-\frac12h^{\rm GW}_{ab}\bar\ell^b_{\not\,0}\,,
\eeq
recalling Eq. \eqref{barellgw}.
Finally, Ref. \cite{Book:2010pf} uses an affine parametrization for the photon 4-momentum \eqref{Kbf}, whereas the solution \eqref{barellgw} for $\bar\ell_h^{{\rm GW}\,a}$ is given in terms of the non-affine parameter $\sigma$ given by Eq. \eqref{sigmasolgw}, so that one should also replace $\sigma$ by $\lambda$ before evaluating at the observer to show the agreement with Eq. (39). 
However, Ref. \cite{Book:2010pf} does not discuss how to implement this model in the case of an actual satellite-based astrometric mission, as we have done before by computing the main satellite observables.

\section{Concluding remarks}

The passage of a gravitational wave is expected to induce a time-dependent periodic shift on the apparent positions of stars, entering the astrometric solution in a characteristic way.
We have computed the corrections to the main observables of satellite-based astrometric missions due to the interaction with a monochromatic plane gravitational wave in the RAMOD framework.
Such corrections turn out to depend on the characteristic parameters of the wave, associated with the amplitudes of the two polarization modes, the direction of propagation and the frequency. 
We have also improved the reference astrometric solution for Gaia by including the effect of the quadrupole moment of the planets, generalizing previous results \cite{Crosta:2017eco}.
The quadrupole contribution to the Gaia observables could be directly implemented in the current data processing.
In contrast, detecting GW effects would require suitable search algorithms, data compressing and optimization techniques reducing the dimensionality of the parameter space. The methods discussed in Refs. \cite{Moore:2017ity,Klioner:2017asb} are very promising, so that we expect that an efficient algorithm will be soon available before the final Gaia data release.

Future satellite-based astrometric missions are planned to reach a significantly improved level of accuracy, so that they are better suited to measure GW effects.
In the meantime, the description of the gravitational field of the Solar System should become more accurate, by including further PN terms in the gravitational potentials as well as by relaxing some simplifying assumption on the multipolar structure and proper motion of the planets valid at the microarcsecond level only \cite{Zschocke:2015mca,Zschocke:2016rng,Zschocke:2016ush}.

\appendix

\section{Light propagation in the field of uniformly moving quadrupolar bodies}

The solution of the photon equations of motion is given by \eqref{solbarlfin}--\eqref{calEsol} in terms of the quantities \eqref{Hdefs}, which are listed below in the case of uniformly moving extended bodies endowed with mass quadrupole moment.
It is enough to show the functions $H(\sigma)$, $H^a(\sigma)$ and ${\mathcal H}^a(\sigma)$, because $H^t(\sigma)=-\delta_{ab}H^a(\sigma){\tilde v}^b$ and ${\mathcal H}^t(\sigma)=-\delta_{ab}{\mathcal H}^a(\sigma){\tilde v}^b$, from the relation $\partial_t h=-{\tilde v}^a\partial_a h$ and the assumed constant value of ${\tilde v}^a$.
We will write such functions in the form
\begin{eqnarray}
H(\sigma)&=&H^{(0)}(\sigma)+\epsilon H^{(1)}(\sigma)+O(2)
\,,\nonumber\\
H^a(\sigma)&=&H^a{}^{(0)}(\sigma)+\epsilon H^a{}^{(1)}(\sigma)+O(2)
\,,\nonumber\\
{\mathcal H}^a(\sigma)&=&{\mathcal H}^a{}^{(0)}(\sigma)+\epsilon {\mathcal H}^a{}^{(1)}(\sigma)+O(2)
\,,
\end{eqnarray}
since our solution is accurate up to the order $O(3)$ included.
A solution for quadrupolar bodies in the RAMOD framework has already been presented in Ref. \cite{Crosta:2015jia}, but is affected by several mistakes, as shown in Appendix B.
Our derivation discussed in Section III is different from that of Ref. \cite{Crosta:2015jia} (a term-by-term comparison is not possible), so we will give below the final solution only for the momentum and orbit of the photon.

For the sake of simplicity we will drop the summation symbol over the bodies in the gravitational potential $h=\sum_{A=1}^N h_A$ as well as the label $A$, thus referring to a single source with potential \eqref{hquaddef}.
Quantities in bold are three dimensional vectors, i.e., $\boldsymbol{a}=a^i\partial_i$, so that both scalar and cross product between them are meant to be the standard operations in an Euclidean space and referred to standard Cartesian coordinates, i.e., $\boldsymbol{a}\cdot\boldsymbol{c}=\delta_{ij}a^ic^j$ and $(\boldsymbol{a}\times\boldsymbol{c})^i=\epsilon_{ijk}a^jc^k$.

We will use the following definitions:

\begin{eqnarray}
b^a&=&[\boldsymbol{\bar\ell}_{\not\,0}\times(\boldsymbol{r}_{\rm obs}\times \boldsymbol{\bar\ell}_{\not\,0})]^a
=r^a_{\rm obs}-\bar\ell^a_{\not\,0}\robsdotl
\,,\nonumber\\
d^a&=&[\boldsymbol{\bar\ell}_{\not\,0}\times(\tilde{\boldsymbol{v}}\times \boldsymbol{\bar\ell}_{\not\,0})]^a
={\tilde v}^a-\bar\ell^a_{\not\,0}\vdotl
\,,
\end{eqnarray}
so that $b^2=\delta_{ab}b^ab^b=r_{\rm obs}^2-\robsdotl^2$, and
\begin{eqnarray}
C_n&=&\frac{1}{r^n}-\frac{1}{r_{\rm obs}^n}\,,\qquad
F_n=\frac{\rdotl}{r^n}-\frac{\robsdotl}{r_{\rm obs}^n}\,, \nonumber\\
S&=&r-r_{\rm obs}-\frac{\robsdotl}{r_{\rm obs}}\sigma\,,
\end{eqnarray}
where
\beq
\rdotl=\robsdotl+\sigma+O(2)\,,
\eeq
so that
\beq
r=\sqrt{\sigma^2+r_{\rm obs}^2+2\sigma\robsdotl}
=\sqrt{b^2+\rdotl^2}\,,
\eeq
and
\begin{eqnarray}
D^a_n&=&d^a-n\frac{b^z}{b^2}\vdotb\,, \nonumber\\
X_n&=&1-n(\bar\ell^z_{\not\,0})^2\,, \qquad
Y=(\bar\ell^z_{\not\,0})^2-\frac{(b^z)^2}{b^2}\,.
\end{eqnarray}

\subsection{Monopole solution}

The correct solution for a uniformly moving mass monopole has already been given in Ref. \cite{Crosta:2017eco}, although not explicitly pointed out there.
We recall it below for completeness:

\begin{eqnarray}
\label{Hvariemono}
H_M^{(0)}(\sigma)&=&
GM\ln\left[\frac{r+\rdotl}{r_{\rm obs}+\robsdotl}\right]
\,, \nonumber\\
H_M^a{}^{(0)}(\sigma)&=&
GM\left(\bar\ell^a_{\not\,0}C_1-\frac{b^a}{b^2}F_1\right)
\,, \nonumber\\
{\mathcal H}_M^a{}^{(0)}(\sigma)&=&
\bar\ell^a_{\not\,0}\left(H_M^{(0)}(\sigma)-\frac{GM}{r_{\rm obs}}\sigma\right)
-GM\frac{b^a}{b^2}S\,,\nonumber\\
\end{eqnarray}
and

\begin{widetext}

\begin{eqnarray}
H_M^{(1)}(\sigma)&=&
\vdotl H_M^{(0)}
+\frac{GM}{r+\rdotl}\left[\frac{r-r_{\rm obs}+\sigma}{r_{\rm obs}+\robsdotl}\vdotb-\vdotl\sigma\right]
\,, \nonumber\\
H_M^a{}^{(1)}(\sigma)&=&
\vdotl H_M^a{}^{(0)}(\sigma)\nonumber\\
&&
+\frac{GM}{b^2}\left\{\frac{r_{\rm obs}}{r}D^a_2S
-b^a\left[\vdotl\robsdotl+\vdotb\right]C_1
+[b^a\vdotl-\bar\ell^a_{\not\,0}\vdotb]F_1
\right\}
\,, \nonumber\\
{\mathcal H}_M^a{}^{(1)}(\sigma)&=&
\vdotl{\mathcal H}_M^a{}^{(0)}(\sigma)
+\bar\ell^a_{\not\,0}H_M^{(1)}(\sigma)-d^aH_M^{(0)}(\sigma)\nonumber\\
&&
-\frac{GM}{b^2}\left\{
\left[b^a\vdotl
+\bar\ell^a_{\not\,0}\vdotb\right]S
-rr_{\rm obs}D^a_2F_1
+\frac{b^a}{r_{\rm obs}}\left[\vdotl\robsdotl-\vdotb\right]\sigma
\right\}\,.
\end{eqnarray}

\subsection{Quadrupole solution}

For a mass quadrupole one gets (see also Ref. \cite{Bini:2014tiu} for the static case)
\begin{eqnarray}
\label{solfinquad}
H_{Q}^{(0)}(\sigma)&=&GM(R^{\rm eq})^2J_2\left\{
\bar\ell^z_{\not\,0}b^zC_3
+\frac12YF_3
+\frac1{2b^2}(X_3+2Y)F_1
\right\}
\,, \nonumber \\
H^a_{Q}{}^{(0)}(\sigma)&=&GM(R^{\rm eq})^2J_2\left\{
b^a\left[
-3\bar\ell^z_{\not\,0}b^zC_5
-\frac32YF_5
-\frac1{2b^2}(X_5+4Y)\left(\frac{2}{b^2}F_1+F_3\right)\right]
\right.\nonumber\\
&&\left.
+\bar\ell^a_{\not\,0}\left[
\frac12X_5C_3
+\frac32b^2YC_5
+\frac{\bar\ell^z_{\not\,0}b^z}{b^2}\left(\frac{2}{b^2}F_1
+F_3
-3b^2F_5\right)
\right]
+\delta^a_z\left[
-\frac{b^z}{b^2}\left(F_3+\frac{2}{b^2}F_1\right)
+\bar\ell^z_{\not\,0}C_3
\right]
\right\}
\,, \nonumber
\end{eqnarray}
\begin{eqnarray}
{\mathcal H}^a_{Q}{}^{(0)}(\sigma)&=&GM(R^{\rm eq})^2J_2\left\{
b^a\left[
\frac12YC_3
-\frac{\bar\ell^z_{\not\,0}b^z}{b^2}\left(\frac{2}{b^2}F_1
+F_3\right)
+\frac1{2b^2}(X_5+4Y)\left(-\frac{2}{b^2}(r-r_{\rm obs})+C_1\right)
\right.\right.\nonumber\\
&&\left.
+\left[
\frac1{2b^2}(X_5+4Y)\left(\frac{2}{b^2}+\frac{1}{r_{\rm obs}^2}\right)\frac{\robsdotl}{r_{\rm obs}}
+\frac{3}{r_{\rm obs}^5}\left(\bar\ell^z_{\not\,0}b^z+\frac12Y\robsdotl\right)
\right]\sigma
\right]\nonumber\\
&&
+\bar\ell^a_{\not\,0}\left[
\frac1{2b^2}(X_5+2Y)F_1
+\frac12YF_3
+\frac{\bar\ell^z_{\not\,0}b^z}{b^2}\left(\frac{2}{b^2}(r-r_{\rm obs})-C_1+b^2C_3\right)
\right.\nonumber\\
&&\left.
-\left[
\frac{1}{2r_{\rm obs}^3}\left(X_5+3\frac{b^2}{r_{\rm obs}^2}Y\right)
+\frac{2\bar\ell^z_{\not\,0}b^z}{b^4}\left(1+\frac{b^2}{2r_{\rm obs}^2}\left(1-\frac{3b^2}{r_{\rm obs}^2}\right)\right)\frac{\robsdotl}{r_{\rm obs}}
\right]
\sigma
\right]\nonumber\\
&&\left.
+\delta^a_z\left[
\frac{b^z}{b^2}\left(C_1
+\frac{\robsdotl}{r_{\rm obs}^3}\sigma
-\frac{2}{b^2}S\right)
+\bar\ell^z_{\not\,0}\left(\frac{1}{b^2}F_1-\frac{1}{r_{\rm obs}^3}\sigma\right)
\right]
\right\}\,,
\end{eqnarray}
and
\begin{eqnarray}
H_{Q}^{(1)}(\sigma)&=&GM(R^{\rm eq})^2J_2\left\{
-\left(\bar\ell^z_{\not\,0}D^z_{-1}+\frac{b^z}{b^2}D^z_1\robsdotl\right)F_3
+\frac{1}{b^2}\left[-\frac{1}{b^2}\robsdotl(2b^zD^z_2+X_1\vdotb)+\bar\ell^z_{\not\,0}D^z_2\right]F_1\right.\nonumber\\
&&
+\left[\frac12Y\vdotb-b^z(D^z_1-\bar\ell^z_{\not\,0}\vdotl)+\left(\frac12Y\vdotl+\bar\ell^z_{\not\,0}d^z\right)\robsdotl\right]C_3
\nonumber\\
&&\left.
+\frac{1}{2b^2}(X_3+2Y)[\vdotl\robsdotl-\vdotb]C_1
\right\}
\,,\nonumber
\end{eqnarray}
\begin{eqnarray}
H^a_{Q}{}^{(1)}(\sigma)&=&GM(R^{\rm eq})^2J_2\left\{
3\left[
b^a\left(\frac{b^z}{b^2}D^z_1\robsdotl+\bar\ell^z_{\not\,0}D^z_{-1}\right)
+\bar\ell^a_{\not\,0}\left(-\bar\ell^z_{\not\,0}d^z\robsdotl+b^zD^z_1-\frac{3}{2}Y\vdotb\right)\right.\right.
\nonumber\\
&&\left.
+d^a\left(-\frac{1}{2}Y\robsdotl+b^z\bar\ell^z_{\not\,0}\right)
\right]
F_5\nonumber\\
&&
+\frac{1}{b^2}\left[
b^a\left(\frac{1}{b^2}(4D^z_2b^z+X_1\vdotb)\robsdotl-\bar\ell^z_{\not\,0}D^z_4\right)
-\bar\ell^a_{\not\,0}\left(D^z_2(b^z-\bar\ell^z_{\not\,0}\robsdotl)+\frac12X_3\vdotb\right)\right.
\nonumber\\
&&\left.
-d^a\left(\frac12(X_5+4Y)\robsdotl+b^z\bar\ell^z_{\not\,0}\right)
-\delta^a_z\left(D^z_2\robsdotl+\bar\ell^z_{\not\,0}\vdotb\right)
\right]
F_3\nonumber\\
&&
+\frac{1}{b^4}\left[
2b^a\left(\frac{2}{b^2}(2b^zD^z_3+X_1\vdotb)\robsdotl-\bar\ell^z_{\not\,0}D^z_4\right)
+\bar\ell^a_{\not\,0}\left(2\bar\ell^z_{\not\,0}D^z_4\robsdotl-2D^z_2b^z-X_3\vdotb\right)\right.
\nonumber\\
&&\left.
-d^a\left[(X_5+4Y)\robsdotl+2b^z\bar\ell^z_{\not\,0}\right]
-2\delta^a_z\left(D^z_4\robsdotl+\bar\ell^z_{\not\,0}\vdotb\right)
\right]
F_1\nonumber\\
&&
+3\left[
b^a\left(-\left(\bar\ell^z_{\not\,0}d^z+\frac12Y\vdotl\right)\robsdotl+b^z(D^z_1-\bar\ell^z_{\not\,0}\vdotl)-\frac12Y\vdotb\right)\right.
\nonumber\\
&&
+\bar\ell^a_{\not\,0}\left(\left(-b^z(D^z_1+\bar\ell^z_{\not\,0}\vdotl)+Y\vdotb\right)\robsdotl-b^2\left(\bar\ell^z_{\not\,0}D^z_{-1}-\frac12Y\vdotl\right)\right)\nonumber\\
&&\left.
-d^a\left(\frac12Yb^2+\bar\ell^z_{\not\,0}b^z\robsdotl\right)
\right]
C_5\nonumber\\
&&
+\left[
\frac{b^a}{2b^2}\left(\vdotb-\vdotl\robsdotl\right)(X_5+4Y)
-\frac12d^aX_5\right.\nonumber\\
&&
+\bar\ell^a_{\not\,0}\left(-\frac{b^z}{b^2}\bar\ell^z_{\not\,0}\left(\vdotb-\vdotl\robsdotl\right)+5\bar\ell^z_{\not\,0}d^z+\frac12X_5\vdotl\right)
\nonumber\\
&&\left.
+\delta^a_z\left(-\frac{b^z}{b^2}\vdotl\robsdotl+\bar\ell^z_{\not\,0}\vdotl-D^z_1\right)
\right]
C_3\nonumber\\
&&\left.
+\frac{1}{b^4}\left(\vdotb-\vdotl\robsdotl\right)\left[b^a(X_5+4Y)-2\bar\ell^a_{\not\,0}b^z\bar\ell^z_{\not\,0}
+2\delta^a_zb^z\right]C_1
\right\}
\,,\nonumber
\end{eqnarray}
\begin{eqnarray}
{\mathcal H}^a_{Q}{}^{(1)}(\sigma)&=&GM(R^{\rm eq})^2J_2\left\{
\left[
\frac{b^a}{b^2}\left(-\bar\ell^z_{\not\,0}D^z_2\robsdotl+b^z(D^z_1-\bar\ell^z_{\not\,0}\vdotl)-Y\vdotb\right)\right.\right.
\nonumber\\
&&\left.
+\bar\ell^a_{\not\,0}\left(-\frac{b^z}{b^2}D^z_1\robsdotl-\bar\ell^z_{\not\,0}D^z_{-2}+\frac12Y\vdotl\right)
+d^a\left(-\frac{b^z}{b^2}\bar\ell^z_{\not\,0}\robsdotl-\frac12Y\right)
\right]
F_3\nonumber\\
&&
+\frac{1}{b^2}\left[
2\frac{b^a}{b^2}\left(-\bar\ell^z_{\not\,0}D^z_4\robsdotl+b^z(D^z_1-\bar\ell^z_{\not\,0}\vdotl)-Y\vdotb\right)\right.
\nonumber\\
&&
+\bar\ell^a_{\not\,0}\left(\frac1{b^2}\left(-2b^zD^z_1-(X_5+2Y)\vdotb\right)\robsdotl+\bar\ell^z_{\not\,0}(4D^z_1-d^z)+\frac12(X_5+2Y)\vdotl\right)
\nonumber\\
&&\left.
-d^a\left(2\frac{b^z}{b^2}\bar\ell^z_{\not\,0}\robsdotl+\frac12(X_5+2Y)\right)
-\delta^a_z\left(d^z+\frac{2}{b^2}\bar\ell^z_{\not\,0}\vdotb\robsdotl-\bar\ell^z_{\not\,0}\vdotl\right)
\right]
F_1\nonumber\\
&&
+\left[
b^a\left(-\frac1{b^2}\left(b^zD^z_1+\bar\ell^z_{\not\,0}b^z\vdotl\right)\robsdotl-\bar\ell^z_{\not\,0}d^z+Y\vdotl\right)\right.
\nonumber\\
&&
+\bar\ell^a_{\not\,0}\left(\left(\bar\ell^z_{\not\,0}d^z+\frac12Y\vdotl\right)\robsdotl-b^z(D^z_1-2\bar\ell^z_{\not\,0}\vdotl)+Y\vdotb\right)\nonumber\\
&&\left.
+d^a\left(-b^z\bar\ell^z_{\not\,0}+\frac12Y\robsdotl\right)
\right]
C_3\nonumber\\
&&
+\frac{1}{b^2}\left[
b^a\left(\frac{1}{b^2}\left(-2b^z(2D^z_1+\bar\ell^z_{\not\,0}\vdotl)-(X_5+4Y)\vdotb\right)\robsdotl+\bar\ell^z_{\not\,0}D^z_2+(X_5+4Y)\vdotl\right)\right.
\nonumber\\
&&
+\bar\ell^a_{\not\,0}\left(\left(-\bar\ell^z_{\not\,0}D^z_2+\frac12(X_5+2Y)\vdotl\right)\robsdotl
+b^z(D^z_1-2\bar\ell^z_{\not\,0}\vdotl)\right)\nonumber\\
&&\left.
+d^a\left(\frac12(X_5+4Y)\robsdotl+\bar\ell^z_{\not\,0}b^z\right)
+\delta^a_z\left((D^z_2+\bar\ell^z_{\not\,0}\vdotl)\robsdotl+2b^z\vdotl\right)
\right]
C_1\nonumber\\
&&
+\frac{1}{b^4}\left[
2b^a\left(\frac{2}{b^2}\left(2b^zD^z_1+(X_5+4Y)\vdotb\right)\robsdotl-\bar\ell^z_{\not\,0}D^z_4-(X_5+4Y)\vdotl\right)\right.
\nonumber\\
&&
+\bar\ell^a_{\not\,0}\left(2\bar\ell^z_{\not\,0}D^z_4\robsdotl-2b^z(d^z-2\bar\ell^z_{\not\,0}\vdotl)-(X_7+4Y)\vdotb\right)\nonumber\\
&&\left.
+d^a\left(-(X_5+4Y)\robsdotl-2b^z\bar\ell^z_{\not\,0}\right)
+\delta^a_z\left(-2D^z_4\robsdotl-2\bar\ell^z_{\not\,0}\vdotb-4b^z\vdotl\right)
\right]
S\nonumber\\
&&
+\left[
\frac{1}{r_{\rm obs}^5}b^a\left(\left(-\frac{r_{\rm obs}^4}{b^4}(X_5+4Y)\vdotl+\frac{r_{\rm obs}^2}{b^2}\left(\bar\ell^z_{\not\,0}D^z_4+\frac12(X_5+4Y)\vdotl\right)-3\frac{b^z}{b^2}\bar\ell^z_{\not\,0}\vdotb\right.\right.\right.\nonumber\\
&&\left.
+\frac32Y\vdotl\right)\robsdotl-2\frac{r_{\rm obs}^4}{b^4}\left(2b^zD^z_1+(X_5+4Y)\vdotb\right)+\frac{r_{\rm obs}^2}{b^2}\left(b^zD^z_1+\frac12(X_5+4Y)\vdotb\right)\nonumber\\
&&\left.
+\frac32Y\vdotb+3\bar\ell^z_{\not\,0}b^z\vdotl\right)
+\frac{1}{r_{\rm obs}^3}d^a\left(\frac{b^z}{b^2}\bar\ell^z_{\not\,0}\robsdotl-\frac12Y+\frac12\frac{r_{\rm obs}^2}{b^2}(X_5+4Y)\right)\nonumber\\
&&
+\frac{1}{r_{\rm obs}^5}\bar\ell^a_{\not\,0}\left(\left(2\frac{r_{\rm obs}^4}{b^4}b^z\bar\ell^z_{\not\,0}\vdotl+\frac{r_{\rm obs}^2}{b^2}\left(\frac12X_3\vdotb+b^z(D^z_2-\bar\ell^z_{\not\,0}\vdotl)\right)+\frac32Y\vdotb\right.\right.\nonumber\\
&&\left.\left.
+3\bar\ell^z_{\not\,0}b^z\vdotl\right)\robsdotl-\frac{r_{\rm obs}^4}{b^2}\bar\ell^z_{\not\,0}D^z_4-r_{\rm obs}^2\left(\bar\ell^z_{\not\,0}D^z_{-1}+\frac12X_5\vdotl\right)-\frac32b^2Y\vdotl+3b^z\bar\ell^z_{\not\,0}\vdotb\right)\nonumber\\
&&\left.\left.
+\frac{1}{r_{\rm obs}^3}\delta^a_z\left(\frac{1}{b^2}\left(\frac{b^z}{b^2}(b^2-2r_{\rm obs}^2)\vdotl+\bar\ell^z_{\not\,0}\vdotb\right)\robsdotl+\frac{r_{\rm obs}^2}{b^2}D^z_4+\frac{b^z}{b^2}\vdotb-\bar\ell^z_{\not\,0}\vdotl\right)
\right]
\sigma
\right\}
\,.\nonumber\\
\end{eqnarray}

\end{widetext}

\section{Correcting formulas in previous RAMOD papers}

The theoretical framework of the RAMOD model is developed in Refs. \cite{deFelice:2004nf,deFelice:2006xm}, whereas Ref. \cite{Crosta:2015jia} contains the main application to the Gaia context.
Ref. \cite{deFelice:2004nf} deals with the static case. The generalization to the dynamical case is discussed in Ref. \cite{deFelice:2006xm}, which however contains an incorrect equation for the component $\bar\ell^0$ of the relative-observer spatial momentum (cf. Eq. (16) there with Eq. \eqref{RME_static1} here). 
Finally, Ref. \cite{Crosta:2015jia} has a number of results requiring correction, which are summarized below.

\begin{enumerate}

\item
The equation governing the evolution of $\bar\ell^0$ along the photon path given in Eq. (16) of Ref. \cite{deFelice:2006xm} (see also Eq. (2) of Ref. \cite{Crosta:2015jia}) is incorrect. It should read instead as in Eq. \eqref{RME_static1} above.
In fact, the second term in the right hand side of Eq. \eqref{RME_static1} can be neglected at the order $O(3)$, so that the disagreement with Eq. (16) of Ref. \cite{deFelice:2006xm} is due to the term $-\frac12\partial_0h_{00}$ there, which is clearly wrong. 
This can be shown simply by taking the solution for $\bar\ell^0$ in terms of that for the spatial components $\bar\ell^a$ given by the first equation of \eqref{l0vsla} (Eq. (14) of Ref. \cite{deFelice:2006xm}).
Differentiating the latter with respect to $\sigma$ and using Eq. \eqref{eqxRME} then immediately gives the evolution equation \eqref{RME_static1} for $\bar\ell^0$, while it does not reproduce Eq. (16) of Ref. \cite{deFelice:2006xm}, even at the order $O(3)$.
All related formulas containing $\bar\ell^0$ (or its explicit solution after the metric is specified) in subsequent RAMOD papers must reflect this change.

\item
The gravitational potentials entering the spacetime metric are written in Ref. \cite{Crosta:2015jia} in terms of the retarded time (see, e.g., Eq. (18) there in the case of uniformly moving pointlike bodies), which is a function of the global coordinates.  
The derivation of the solution to the photon equations of motion then proceeds without specifying this relation, generating unnecessary additional terms. The final solution for the photon trajectory still contains retarded quantities, so that it is never explicit.
It should necessarily be further transformed using the relation $t_{\rm ret}=t_{\rm ret}(t, x^i)$ and re-expressed in terms of global coordinates, but this is not the case in Ref. \cite{Crosta:2015jia}.

\item
The authors claim that the congruence of static observers is vorticity-free if the sources move with constant velocity (see the sentence after Eq. (57) in Ref. \cite{Crosta:2015jia}). However, this is not the case.
In fact, the components of the vorticity tensor turn out to be $\omega(u)_{ij}=-h_{0[i,j]}$ (see Eq. \eqref{mkinq}, where $h_{0i}=-4\epsilon^3\sum_Ah_A(t,x^i){\tilde v}_A^i$ with potential \eqref{potdef} of the $A$-th source), which vanish only if ${\tilde v}_A^i\equiv0$, since $h_A(t,x^i)$ and their spatial derivatives are nonzero.

In order to have a vorticity-free congruence one should select as fiducial observers a different family of observers, e.g., those with 4-velocity vector $n$ orthogonal to the $t=$constant hypersurfaces (see, e.g., Ref. \cite{Jantzen:1992rg}).
To $O(h)$ we find 
\beq
n=\left(1+\frac12 h_{00}\right)\partial_0-h_{0i}\partial_i\,,
\eeq
with associated kinematical fields
\begin{eqnarray}
a(n)^i&=&-\frac12 h_{00,i}\,,\nonumber\\
\theta(n)_{ij}&=&\frac12 h_{ij,0}-h_{0(i,j)}\,,\nonumber\\
\omega(n)_{ij}&=&0\,.
\end{eqnarray}
Equations \eqref{RME_gen} referred to $n$ then imply

\begin{widetext}

\begin{eqnarray}
\label{RME_zamo}
\frac{\rmd\bar \ell(n)^0}{\rmd \sigma}&=&0\,,\nonumber\\
\frac{\rmd\bar \ell(n)^k}{\rmd \sigma}&=&-\bar \ell(n)^i\bar \ell(n)^j \left(h_{ki,j} - \frac12h_{ij,k}\right)
-\bar \ell(n)^i(h_{ki,0}-h_{0i,k})
+\frac12h_{00,k}
\nonumber\\
&&
+\bar \ell(n)^k\left[\bar \ell(n)^i\bar \ell(n)^j \left(\frac12 h_{ij,0} - h_{0i,j}\right)-\frac12\bar \ell(n)^i h_{00,i}\right]\,.
\end{eqnarray}

\end{widetext}

\item
The solution for the photon trajectory is wrong in the general case, i.e., for moving bodies. 
Moreover, it is also incorrect in the static case. 
In fact, the integration of the photon equations is not carried out correctly.
Consider, for instance, Eq. (72) in Ref. \cite{Crosta:2015jia}, which is correct.
Further integrating this equation would lead to Eq. (127) there, which however has two missing terms, proportional to the integration interval.
Similar mistakes have propagated throughout the paper.

\end{enumerate}

The correct solution for uniformly moving extended bodies endowed with a mass quadrupole moment is given in Appendix A above.

\section*{Acknowledgments}

The authors thank Prof. F. de Felice for useful discussions on the RAMOD model over the years.
A.~G. also thanks M.~T.~Crosta, M.~G.~Lattanzi and A.~Vecchiato for directing his attention to the main literature developing the RAMOD approach during his past research fellowship at the Astrophysical Observatory of Turin about two years ago. 
In fact, careful checking of previous results on the solution of light propagation in the gravitational field of the Solar System led to the need to correct several formulas and redo also the application to quadrupolar bodies from a slightly different perspective, more suitable for the computation of astrometric observables.

\end{document}